\documentclass[prl,twocolumn,superscriptaddress,amsmath,amssymb]{revtex4}

\usepackage{graphicx}
\usepackage{bm}

\newcommand {\Fig}[1] {Figure~\ref{#1}}

\newcommand{\beq}{\begin{equation}}
\newcommand{\eeq}{\end{equation}}

\newcommand{\beqa}{\begin{eqnarray}}
\newcommand{\eeqa}{\end{eqnarray}}

\newcommand{\ket}[1]{\left| #1 \right\rangle}
\newcommand{\bra}[1]{\left\langle #1 \right|}

\newcommand{\JCP}{J. Chem. Phys.}

\newcommand{\JPCM}{J. Phys. Cond. Matt.}

\newcommand{\PTRSA}{Phil.~Trans.~R.~Soc. A}

\newcommand{\PRA}{Phys. Rev. A}
\newcommand{\PRB}{Phys. Rev. B}
\newcommand{\PRL}{Phys. Rev. Lett.}

\newcommand{\RSI}{Rev. Sci. Inst.}

\newcommand{\tonen}{$T_{\rm{1n}}$}
\newcommand{\tonee}{$T_{\rm{1e}}$}
\newcommand{\ttwon}{$T_{\rm{2n}}$}
\newcommand{\ttwoe}{$T_{\rm{2e}}$}
\newcommand{\tstore}{$T_{\rm{store}}$}

\begin{document}

\title{Solid state quantum memory using the $^{31}$P nuclear spin}

\author{John~J.~L.~Morton}
\email{john.morton@materials.ox.ac.uk} \affiliation{Department of Materials, Oxford University, Oxford OX1 3PH, United Kingdom}
\affiliation{Clarendon Laboratory,
Department of Physics, Oxford University, Oxford OX1 3PU, United Kingdom}

\author{Alexei~M.~Tyryshkin}
\affiliation{Department of Electrical Engineering, Princeton
University, Princeton, NJ 08544, USA}

\author{Richard M. Brown}
\affiliation{Department of Materials, Oxford University, Oxford OX1 3PH, United Kingdom}

\author{Shyam~Shankar}
\affiliation{Department of Electrical Engineering, Princeton
University, Princeton, NJ 08544, USA}

\author{Brendon W. Lovett}
\affiliation{Department of Materials, Oxford University, Oxford OX1 3PH, United Kingdom}

\author{Arzhang Ardavan}
\affiliation{Clarendon Laboratory, Department of Physics, Oxford University, Oxford OX1 3PU, United Kingdom}

\author{Thomas Schenkel}
\affiliation{Lawrence Berkeley National Laboratory, 1 Cyclotron Road, Berkeley CA 94720, USA}

\author{Eugene E. Haller}
\affiliation{Lawrence Berkeley National Laboratory, 1 Cyclotron Road, Berkeley CA 94720, USA}
\affiliation{Materials Science Department, University of California, Berkeley, USA}

\author{Joel W. Ager}
\affiliation{Lawrence Berkeley National Laboratory, 1 Cyclotron Road, Berkeley CA 94720, USA}

\author{S.~A.~Lyon}
\affiliation{Department of Electrical Engineering, Princeton University, Princeton, NJ 08544, USA}

\date{\today}
\maketitle

The transfer of information between different physical forms is a central theme in communication and computation, for example between processing entities and memory. Nowhere is this more crucial than in quantum computation~\cite{deutsch85}, where great effort must be taken to protect the integrity of a fragile quantum bit (qubit)~\cite{steane99}.  However, transfer of quantum information is particularly challenging, as the process must remain coherent at all times to preserve the quantum nature of the information~\cite{julsgaard04}. Here we demonstrate the coherent transfer of a superposition state in an electron spin `processing' qubit to a nuclear spin `memory' qubit, using a combination of microwave and radiofrequency pulses applied to $^{31}$P donors in an isotopically pure $^{28}$Si crystal~\cite{kane98,tyr06}.  The state is left in the nuclear spin on a timescale that is long compared with the electron decoherence time and then coherently transferred back to the electron spin, thus demonstrating the $^{31}$P nuclear spin as a solid-state \emph{quantum memory}.  The overall store/readout fidelity is about 90$\%$, attributed to imperfect rotations which can be improved through the use of composite pulses~\cite{mortonbb1}. The coherence lifetime of the quantum memory element at 5.5~K exceeds one second.

Classically, transfer of information can include a copying step, facilitating the identification and correction of errors. However, the no-cloning theorem limits the ability to faithfully copy quantum states across different degrees of freedom~\cite{wootters82}; thus error correction becomes more challenging than for classical information and the transfer of information must take place directly. Experimental demonstrations of such transfer include moving a trapped ion qubit in and out of a decoherence-free subspace for storage purposes~\cite{kielpinski01} and optical measurements of NV centres in diamond~\cite{dutt07}. 

Nuclear spins are known to benefit from long coherence times compared to electron spins, but are slow to manipulate and suffer from weak thermal polarisation.  A powerful model for quantum computation is thus one in which electron spins are used for processing and readout while nuclear spins are used for storage. The storage element can be a single, well-defined nuclear spin, or perhaps a bath of nearby nuclear spins~\cite{dobrovitski06}. $^{31}$P donors in silicon provide an ideal combination of long-lived \mbox{spin-1/2} electron~\cite{tyr03}
and nuclear spins~\cite{mendor}, with the additional advantage of integration with existing technologies~\cite{kane98} and the possibility of single spin detection by electrical measurement~\cite{boehme06, mccamey06, lo07}. Direct measurement of the $^{31}$P nuclear spin by NMR has only been possible at very high doping levels (e.g.~near the metal insulator transition~\cite{hirsch86}). Instead, electron-nuclear double resonance (ENDOR) can be used to excite both the electron and nuclear spin associated with the donor site, and measure the nuclear spin via the electron~\cite{feher59}. This was recently used to measure the nuclear spin-lattice relaxation time \tonen, which was found to follow the electron relaxation time \tonee~over the range 6 to 12~K with the relationship \tonen$ \approx 250$\tonee~\cite{mendor, tyr06}. The suitability of the nuclear spin as a quantum memory element depends more critically on the nuclear coherence time \ttwon, the measurement of which has now been made possible through the storage procedure described here: by varying the storage time and observing the amplitude of the recovered electron coherence. 

\begin{figure}[t] \centerline
{\includegraphics[width=3in]{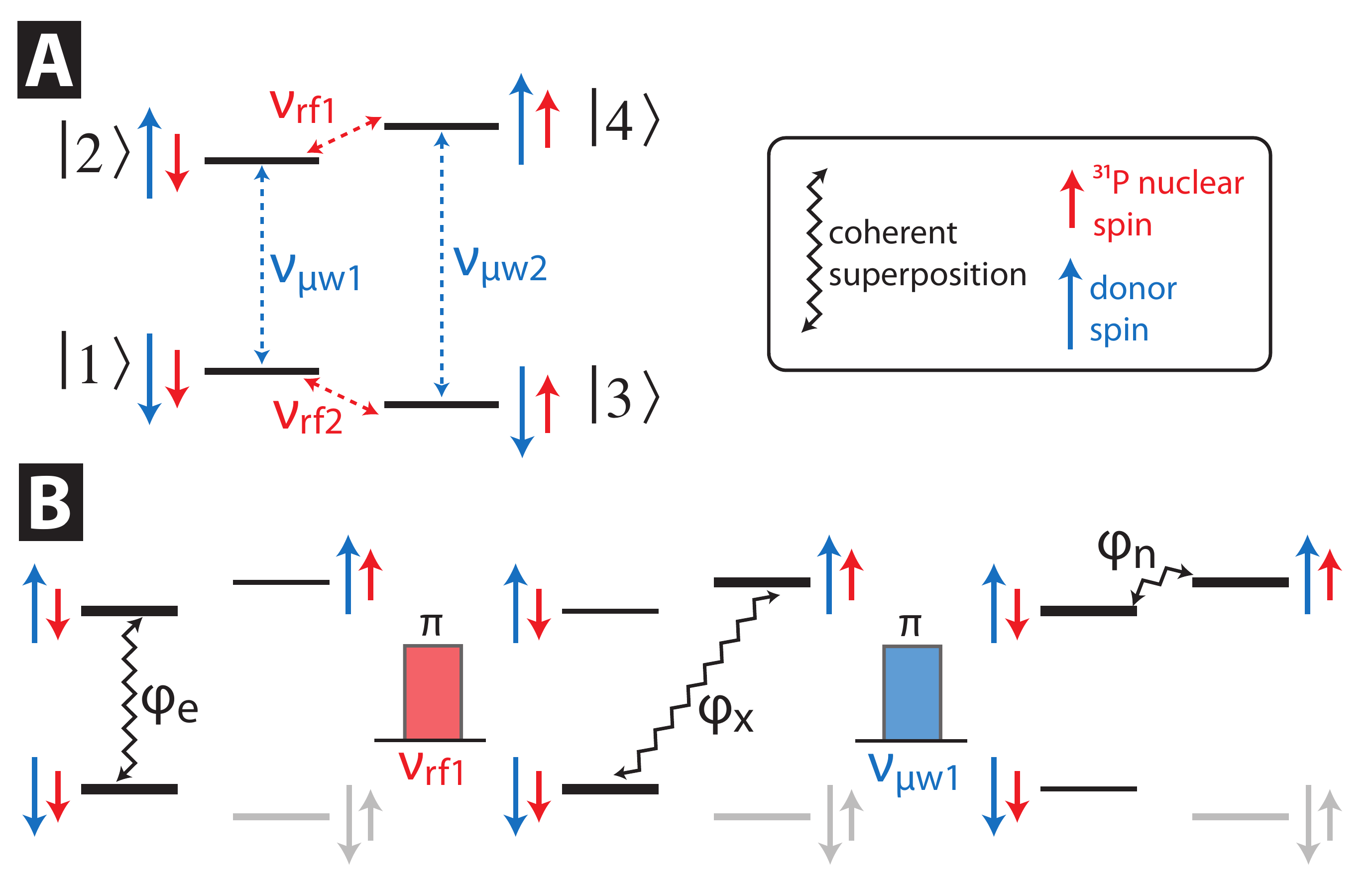}} \caption{\textbf{The level structure of the coupled electron and nuclear spins and scheme for the transfer of a logical qubit within the two physical spin qubits.} (A) The four level system may be manipulated by resonant microwave and radiofrequency (rf) radiation. In our experiments the logical electron spin `processing' qubit is represented by states $\ket{1}$ and $\ket{2}$, whose state can be transferred to a nuclear spin `memory' qubit represented by states $\ket{2}$ and $\ket{4}$. State $\ket{3}$ is never addressed at any point and can be ignored. (B) An electron spin coherence between states $\ket{1}$ and $\ket{2}$ is transferred to the nuclear spin qubit by an rf $\pi$ pulse followed by a microwave $\pi$ pulse. Both pulses must fully excite the transition, and be short compared with the electron and nuclear coherence times. The reverse process is used to transfer the nuclear coherence back to the electron.} \label{fig:one}
\end{figure}

\Fig{fig:one}(B) shows the coherence transfer scheme used for the \emph{write} process from a processing qubit represented by an electron spin degree of freedom, to a memory qubit residing in a nuclear spin degree of freedom. Each $\pi$ pulse is equivalent to a controlled-NOT gate~\cite{mehring03} (with some additional phase which can be ignored) such that the pair of $\pi$ pulses constitute a SWAP gate. The scheme assumes that all pulses are on-resonance and have sufficient bandwidth to completely excite an individual transition. A \emph{read} operation is performed by applying the reverse sequence to bring the coherent state back to the electron spin qubit. Although the phase relationship betweeen the microwave and rf pulses must be constant throughout this process, any phase difference is cancelled out over the course of the write-read process. In practice, this means the microwave and rf sources need not be phase-locked, but must have high phase stability. This is illustrated in calculations following the evolution of the density matrix, provided in the Supplementary Material. 

Although the electron spin qubit can be prepared in a state of high purity using experimentally accessible magnetic fields and temperatures, the small nuclear Zeeman energy results in the nuclear spin being initially in a highly mixed thermal state. However, for the purposes of this quantum memory scheme it is not necessary to perform any pre-cooling of the nuclear spin resource~\footnote{The \emph{write} process from electron to nucleus leaves the nuclear spin qubit in a state of equal purity to the initial electron state. The electron is left in a mixed state, though it is expected to decohere in any case during the storage period.}.

\begin{figure}[t] \centerline
{\includegraphics[width=3.4in]{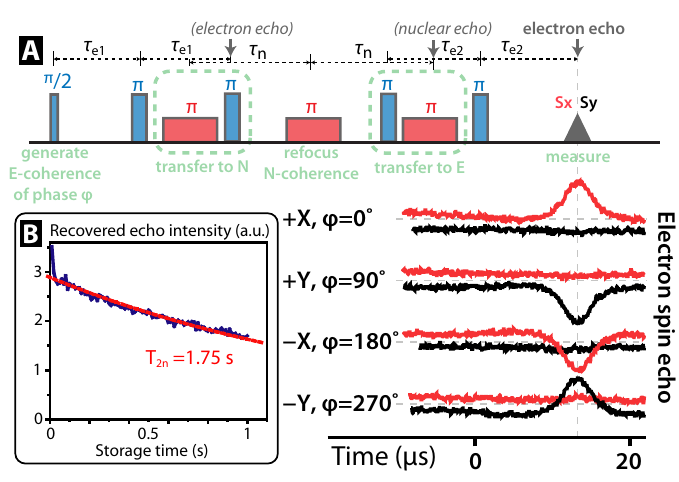}} 
\caption{\textbf{Coherent storage of an electron spin state in a nuclear spin state, using $^{31}$P-doped $^{28}$Si-enriched silicon single crystal}. A) An electron spin coherence is stored in the nuclear spin for 2$\tau_n \approx$ 50~ms, at 7.2~K. The recovered electron spin echo is of comparable intensity to that obtained at the beginning of the sequence, even though the electron spin coherence time \ttwoe~here is about 5~ms. The lifetime of the stored state is limited instead by the nuclear decoherence time \ttwon, which can be measured directly by varying $\tau_n$. B) The recovered echo intensity was measured a function of the storage time at 5.5~K while applying a dynamic decoupling sequence (CPMG) to the nuclear spin, yielding a \ttwon~exceeding 1 second.} \label{fig:store-recover}
\end{figure}

The above model is sufficient given a single electron-nuclear spin pair, or a homogenous ensemble. However, in the experiment described here, we must consider the effects of inhomogeneous broadening across the ensemble of spins being manipulated. The effect of inhomogeneous broadening is to leave some electron (nuclear) spins detuned from the applied microwave (rf) radiation, by $\delta_e$ ($\delta_n$). In a suitable rotating reference frame, electron (nuclear) spin coherence will thus acquire an additional phase at a rate $\delta_e$ ($\delta_n$), while double quantum coherences will acquire phase at a rate $\delta_e$ + $\delta_n$. Thus, inhomogeneous broadening requires the application of carefully placed refocusing pulses to bring all spin packets into focus at key points during the transfer process. In the experiment described here, $\pi/\delta_e \sim$ 2~$\mu$s and $\pi/\delta_n \sim$ 100~$\mu$s.

\Fig{fig:store-recover} shows the practical implementation of a protocol that generates a coherent electron spin state, stores it in a state of the nuclear spin for some time, and then recovers it to the electron state for readout again. The coherence is first generated by a microwave $\pi/2$ pulse of a chosen phase $\varphi$, representing our bit of quantum information. A free induction decay (FID), the \emph{reversible} dephasing of the ensemble, follows this pulse. We apply a refocusing microwave $\pi$ pulse at time $\tau_e$ to initiate a revival in the electron spin coherence. The subsequent rf $\pi$ pulse transfers the coherence from the electron to a double quantum coherence of entangled electron-nuclear spin states. During this period the phase $\delta_e \tau_e$, acquired before the microwave refocusing pulse, continues to unwind so that when the final step of the transfer, a microwave $\pi$ pulse, is applied the effect of the inhomogeneous electron spin packets has been completely refocused. The quantum information that was generated by the first microwave $\pi/2$ pulse now resides entirely in the state of the nucleus. 

This information may be stored in the nuclear state for some extended period so the effects of inhomogeneities on the phase of the nuclear state become appreciable 
and a preparatory rf refocusing pulse must be applied before the information can be recovered. During the nuclear spin echo, the coherence is transferred back to the electron state with a microwave $\pi$ pulse followed by an rf $\pi$ pulse. We apply one further microwave $\pi$ pulse to stimulate an electron spin echo representing the readout event. The lower right panel of \Fig{fig:store-recover} shows the real (red) and imaginary (black) parts of this echo for different initial phases $\varphi$, demonstrating that the encoded phase is recovered through the storage--recovery process, as required for an effective quantum memory element.

The storage time is limited only by the nuclear decoherence time \ttwon, which is in turn limited to 2\tonee~when there is a significant hyperfine interaction ($A \gg 1/$\tonee) between the electron and nuclear spin and in the low-field/high-temperature limit  (see Supplementary Material); \tonee~becomes very long (e.g.\ hours) at low temperatures~\cite{feher59}. A direct measurement of \ttwon~in anything other than highly-doped Si:P has been impossible by traditional NMR means, but our write/read procedure provides a method for performing this measurement by increasing the storage time \tstore~and observing the resulting decay in the recovered electron coherence. \ttwon~obtained in this way indeed follows 2\tonee~approximately over the range 9 to 12~K as expected, though at lower temperatures an additional \ttwon~process appears to play a role, yielding a limit of about 65~ms. A leading candidate for this additional process is slowly fluctuating fields, the effect of which may be mitigated by dynamically decoupling the system~\cite{viola98,bangbang}. By applying a Carr-Purcell-Meiboom-Gill (CPMG) decoupling sequence~\cite{meiboomgill} at a 1~kHz repetition rate to the nuclear spin during the storage period, we were able to obtain much longer decoherence times than for a simple Hahn echo measurement, rising to 1.75 seconds at 5.5~K, as shown in \Fig{fig:store-recover}B. 

Under optimised conditions, \ttwoe~is limited only by magnetic dipole-dipole interactions, and values between 4 and 6.5~ms have been measured in the samples used here, varying according to the donor spin concentration~\cite{tyr03}. Using the nuclear degree of freedom, we have achieved storage times several orders of magnitude longer than \ttwoe. 

\begin{figure}[t] \centerline
{\includegraphics[width=3.2in]{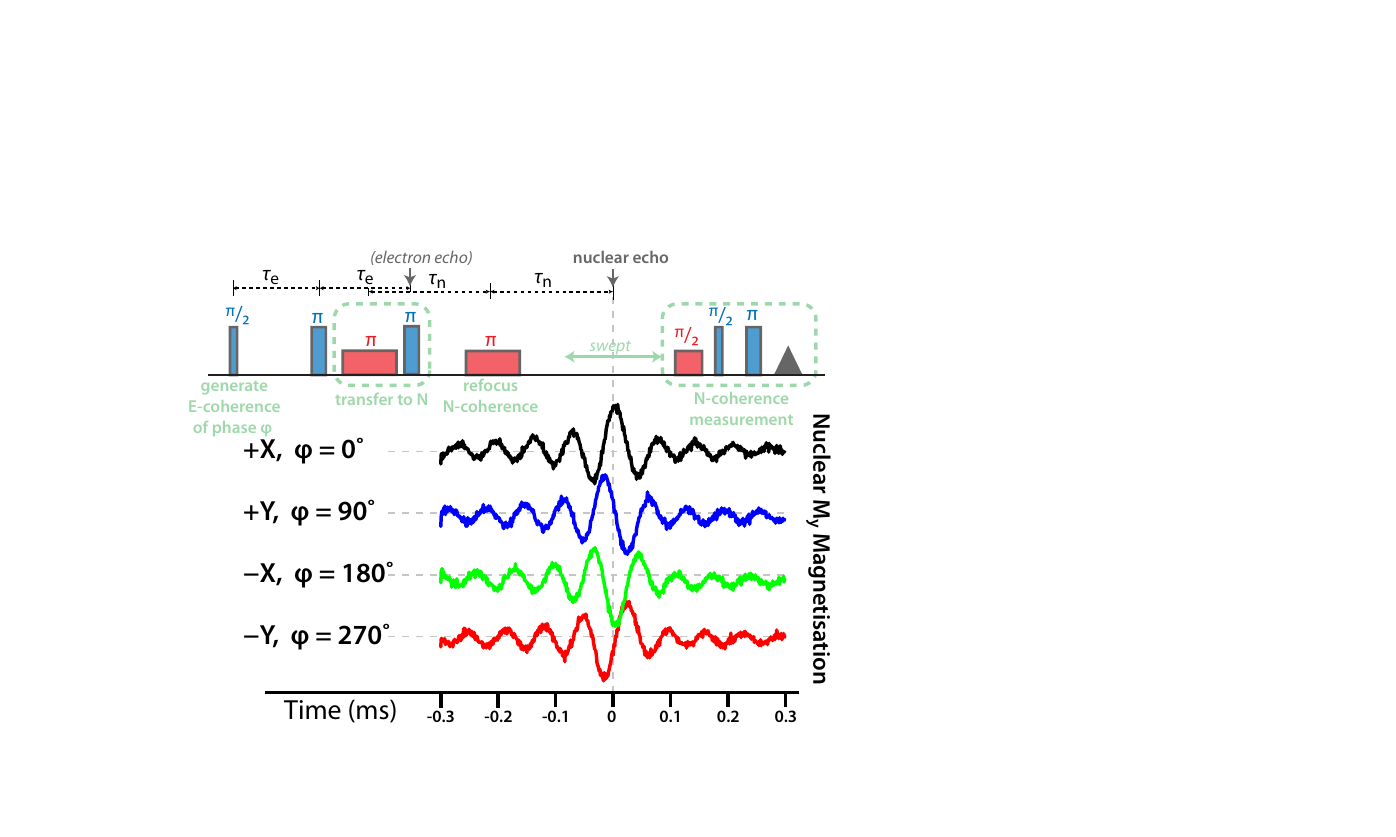}} \caption{\textbf{Observing the nuclear spin coherence during the storage process}. The phase of the initial electron superposition state is determined by the phase of initial $\pi/2$ microwave excitation pulse, which we can control. This state is then transferred to the nuclear spin using the scheme outlined in \Fig{fig:one}. The nuclear spin coherence is read using a process similar to a Ramsey fringe measurement: an rf $\pi/2$ pulse converts nuclear coherence to nuclear polarisation, which is then detected via an electron spin echo measurement selective to one nuclear spin state. The correlation of the phase of the nuclear spin echo to the phase of the original electron spin superposition confirms the coherent nature of the transfer for electron to nuclear spin.} \label{fig:introspect}
\end{figure}


The removal, or substantial detuning of any of the rf pulses in the sequence destroys the recovered echo, confirming the importance of the transfer to the nuclear spin and providing evidence that the stored quantum information does indeed reside in the nuclear state. To go further we require a tool permitting introspection of the state of the nuclear spin during the storage period. 
We therefore applied a sequence to (destructively) probe the nuclear coherence via the electron state, as shown in the upper panel of \Fig{fig:introspect}. The early part of the sequence is as described above: an electron spin coherence is stored in the state of the nucleus. When we would like to observe the state of the nucleus, we apply an rf $\pi/2$ pulse to convert the nuclear coherence into a nuclear polarisation (in the spirit of a Ramsey fringe experiment). A short electron spin echo sequence, selective in one nuclear subspace, then reveals the population of the nuclear level. 

This sequence can be performed at any time; the lower panel of \Fig{fig:introspect} shows the result of observing the state of the nucleus at a range of times for different starting phases $\varphi$, revealing the nuclear spin echo following the rf refocusing pulse. The centre of the rf frequency was intentionally moved off-resonance to produce oscillations in the nuclear echo to aid the identification of the phase of the nuclear coherence. The fact that the phase of the nuclear spin echo follows the phase of the original microwave $\pi/2$ pulse confirms that the information transfer process has remained coherent \footnote{In contrast to other nuclear spin echoes observed via electron-nuclear double resonance (ENDOR)~\cite{hofer86, nc60nuc}, in which an electron spin polarisation is used to create a nuclear spin polarisation, and then a nuclear coherence, this echo represents a coherent state of the electron which has been directly transferred to the nuclear spin.}. 

\begin{figure}[t] \centerline
{\includegraphics[width=3.4in]{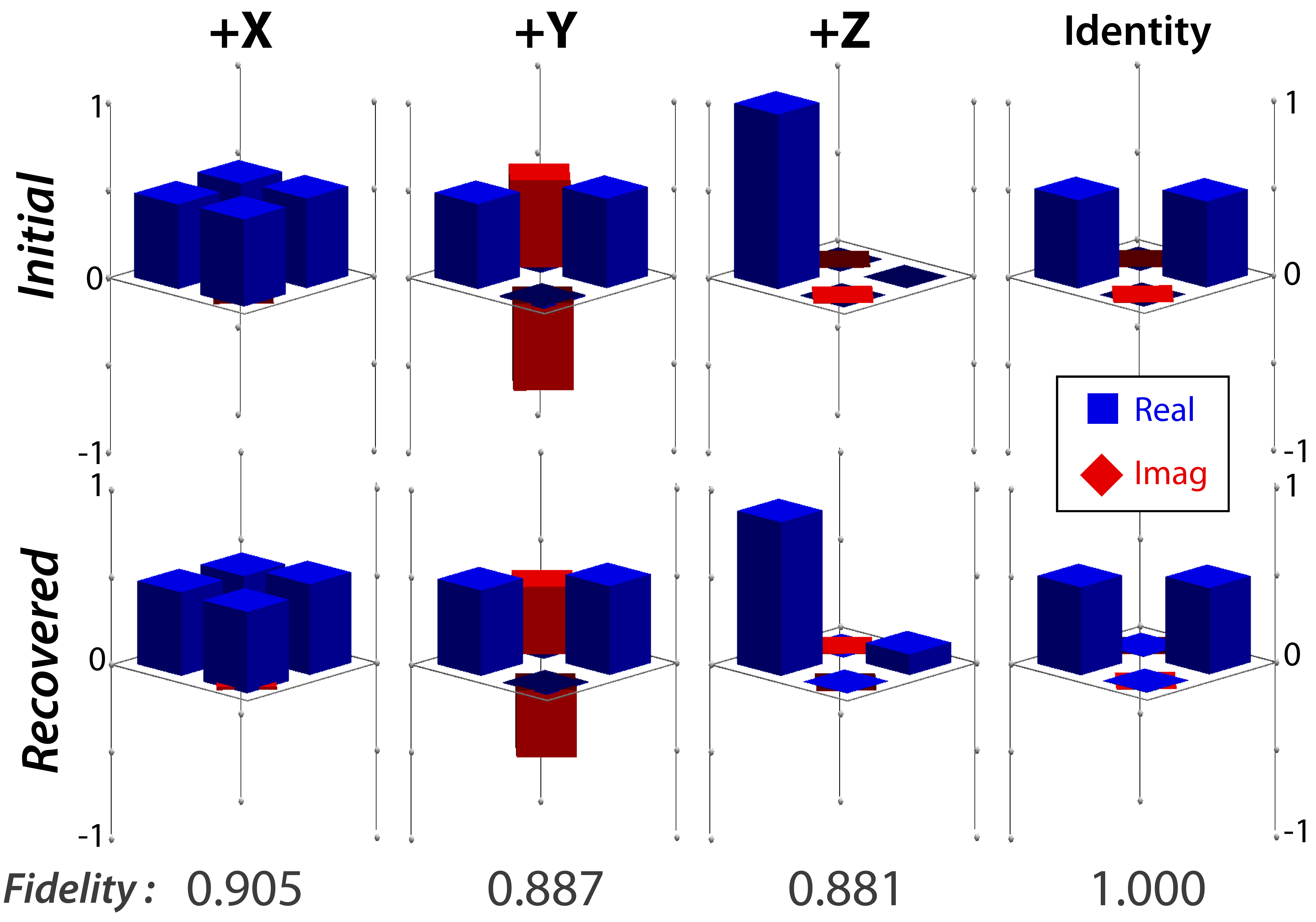}} 
\caption{\textbf{Density matrix tomography for original and recovered states.} Pseudopure states +X, +Y and +Z, and the Identity were prepared in the electron spin qubit and measured (first row). These states were then stored in the nuclear spin degree of freedom and then returned to the electron spin and measured (second row). Tomography was performed by measuring the qubit in the $\sigma_x$, $\sigma_y$ and $\sigma_z$ bases. The fidelity of the quantum memory was obtained by comparing the initial and recovered density matrices.} \label{fig:tomo}
\end{figure}

To demonstrate the generality of the storage sequence described here, we applied it to a wider set of initial states, in particular the $\pm X$, $\pm Y$, $\pm Z$ and Identity basis states and performed density matrix tomography by comparing the original states with those recovered after the \emph{write-read} process (see Supporting Information for full details). The results are summarised in \Fig{fig:tomo}, and show fidelities of approximately 0.90, where the fidelity between initial (pseudo) pure state $\rho_0$ and recovered state $\rho_1$ is defined as $F = \bra{\psi}\rho_1\ket{\psi}$, where $\rho_0 = \ket{\psi}\bra{\psi}$.
We attribute the reduced fidelity to a $\sim5\%$ error in each of the seven microwave and rf pulses applied over the course of the sequence, which is entirely consistent with previous measurements of pulse fidelities~\cite{morton04}. Such errors are mostly systematic, and may be corrected through the application of composite pulses, as previously demonstrated in both EPR and NMR~\cite{cummins03, mortonbb1}. By replacing some of the microwave pulses with BB1 composite pulses we were able to improve the overall fidelity to approximately 0.97 and further improvements are to be expected with greater control of the rf pulse phases.

As the experimental challenges of quantum information processing have become better understood, the importance of hybrid quantum systems in models for quantum information has emerged~\cite{taylor05,thaker06,rabl06,cory07}. The approach described here demonstrates the advantages of such hierarchical models and has a broad applicability in systems where there is a substantial asymmetry in relaxation times. Storage can be driven globally, as shown here, or locally, using EPR gates~\cite{vandersypen-localgate} or Stark tuning~\cite{kane98}. Furthermore, our protocol for faithfully transferring a coherent electron spin state to the nuclear spin offers a route to projective measurements of the qubit state through proposed spectrally sensitive single-spin-detection methodologies such as STM-detected or electrically-detected EPR~\cite{sarovar07}.

We thank Andrew Briggs for comments and support and Ralph Weber and Bruker Biospin for instrumentational support.  We thank Penn Weaver of ASiMI for zone-refining and H. Riemann of the Institut f\"ur Kristallz\"uchtung for float zone processing of the  $^{28}$Si crystals used in this work. This research is supported by the National Security Agency (MOD 713106A) and the EPSRC through the QIP IRC www.qipirc.org (GR/S82176/01) and CAESR (EP/D048559/1). JJLM is supported by St. John's College, Oxford. AA and BWL are supported by the Royal Society. Work at Princeton received support from the NSF through the Princeton MRSEC (DMR-0213706).  Work at LBNL was supported by the Director, Office of Science, Office of Basic Energy Sciences, Materials Sciences and Engineering Division, of the U.S. Department of Energy (DE-AC02-05CH11231).

\clearpage
\newpage

\begin{widetext}
\section{Methods}

Si:P consists of an electron spin S=1/2 (g = 1.9987) coupled to the  
nuclear spin $I=1/2$ of $^{31}$P through a hyperfine coupling  
$A=117$~MHz~\cite{
feher59}, and is described by an isotropic  
spin Hamiltonian (in angular frequency units):
\begin{equation}\label{Hamiltonian}
\mathcal{H}_0=\omega_e S_z - \omega_I I_z + A \!\cdot\! \vec{S} \!\cdot\!
\vec{I},
\end{equation}
where $\omega_e=g\beta B_0/\hbar$ and $\omega_I=g_I\beta_n B_0/\hbar$ are  
the electron and nuclear Zeeman frequencies, $g$ and $g_I$ are the  
electron and nuclear g-factors, $\beta$ and $\beta_n$ are the Bohr and  
nuclear magnetons, $\hbar$ is Planck's constant and $B_0$ is the magnetic  
field applied along $z$-axis in the laboratory frame. The X-band EPR  
signal comprises two lines (one for each nuclear spin projection $M_I =  
\pm 1/2$). Our experiments were performed on the high-field line of the  
EPR doublet corresponding to $M_I=-1/2$.

Single crystal samples were used, as epilayers of $^{28}$Si have a biaxial  
residual stress that broadens the $^{31}$P ENDOR line and makes it  
difficult to fully excite. $^{28}$Si-enriched single crystals with a  
residual $^{29}$Si concentration of 800 ppm were produced by decomposing  
isotopically enriched silane in a recirculating reactor to produce poly-Si  
rods, followed by floating zone crystallization~\cite{ager03}.  To reduce  
spin-spin coupling effects, the phosphorus concentration was reduced from  
an initial value of near $1\cdot10^{15}$ cm$^{-3}$ to 2--5$\cdot10^{14}$  
cm$^{-3}$ by five passes of zone refining followed by floating zone  
crystallisation.

Pulsed EPR experiments were performed using an X-band (9-10~GHz) Bruker  
EPR spectrometer (Elexsys 580) equipped with a low temperature helium-flow  
cryostat (Oxford CF935). The temperature was controlled with a precision  
greater than $0.05$~K using calibrated temperature sensors (Lakeshore  
Cernox CX-1050-SD) and an Oxford ITC503 temperature controller.

For most measurements, microwave pulses for $\pi$/2 and $\pi$ rotations of  
the electron spin were set to 700 and 1400~ns, and no travelling wave tube  
(TWT) amplifier was used. For CPMG and BB1 experiments, an Amplifier  
Research 20W solid state CW amplifier was used, with $\pi/2$ and $\pi$  
pulses 80 and 160~ns respectively. RF pulses of 20~$\mu$s were used for  
$\pi$ rotations of the $^{31}$P nuclear spins. During CPMG, up to 1000 refocusing pulses
were applied during a single sequence.

\section{Supplementary Information}

\subsection{A. Effect of radiofrequency and microwave phases}
The chosen basis is: \beq
(S,I)=\left[\left(\frac{1}{2},\frac{1}{2}\right),  
\left(-\frac{1}{2},\frac{1}{2}\right),\left(\frac{1}{2},-\frac{1}{2}\right),
\left(-\frac{1}{2},-\frac{1}{2}\right)\right],
\eeq
where $S$ represents the electron donor spin, and $I$ the $^{31}$P nuclear  
spin. All pulses are assumed to be selective on a particular electron or  
nuclear spin transition, as illustrated in \Fig{fig:one}. The phase of the  
initial $\pi/2$ microwave pulse, $\varphi_{e}$, determines the phase of the  
initial electron spin coherence. All other microwave pulses have phase  
$\varphi_{mw}$, while that of the rf pulses is $\varphi_{rf}$.

The initial spin density matrix, neglecting any nuclear spin polarisation,  
is proportional to $(\mathbb{I}+\beta S_z)$, where $\mathbb{I}$ is the  
Identity matrix and $\beta=-\frac{g \mu_B B_0}{kT}$. This can be rewritten in pseudopure state form, neglecting the majority of the Identity component and omitting the constant factor $\beta$:
\beq \rho_0=  (S_z+\mathbb{I}/2)/2 = \rho_{th}=
\left(%
\begin{array}{cccc}
  1/2  & 0 & 0 & 0 \\
   0 & 0 & 0 & 0  \\
   0 & 0 & 1/2 & 0 \\
   0 & 0 & 0 & 0  \\
\end{array}%
\right). \eeq

After the initial (coherence-generating) $\pi/2$ microwave pulse:
\beq \rho_1=
\left(%
\begin{array}{cccc}
   1/4 & \exp{(-i\varphi_e)}/4 & 0 & 0 \\
   \exp{(i\varphi_e)}/4 & 1/4 & 0 & 0  \\
   0 & 0 & 1/2 & 0 \\
   0 & 0 & 0 & 0  \\
\end{array}%
\right). \eeq

The next two pulses, $\pi_{RF}$ followed by $\pi_{mw}$, transfer this  
coherence to the nuclear spin:
\beq \rho_2=
\left(%
\begin{array}{cccc}
   1/4 & 0 &  \exp{(i(\varphi_e-\varphi_{RF}-\varphi_{mw}))}/4 & 0 \\
   0 & 1/2 & 0 & 0  \\
    \exp{(-i(\varphi_e-\varphi_{RF}-\varphi_{mw}))}/4 & 0 & 1/4 & 0 \\
   0 & 0 & 0 & 0  \\
\end{array}%
\right). \eeq

The coherences here decay with characteristic time \ttwon, which is  
typically much longer than \ttwoe. Upon applying the reverse of the  
transfer sequence above ($\pi_{mw}$ followed by $\pi_{rf}$), the electron  
coherence is revived:

\beq \rho_3= \rho_1 =
\left(%
\begin{array}{cccc}
   1/4 & \exp{(-i\varphi_e)}/4 & 0 & 0 \\
   \exp{(i\varphi_e)}/4 & 1/4 & 0 & 0  \\
   0 & 0 & 1/2 & 0 \\
   0 & 0 & 0 & 0  \\
\end{array}%
\right). \eeq

Thus the relative phase of the microwave and rf sources is cancelled out,  
though both must remain stable over the course of the experiment.

\subsection{B. Electron and nuclear spin relaxation}


Electron relaxation of the system can be modeled using a standard master equation in Lindblad form. In order to represent processes that take the system to thermal equilibrium, both raising and lowering terms are included.
\beq
\dot{\rho}=-\frac{\gamma}{2}(\rho S^{-}S^{+} +S^{-}S^{+}\rho -2S^{+} S^{-})-\frac{\gamma e^{-\beta}}{2}(\rho S^{+}S^{-} +S^{+}S^{-}\rho -2S^{-} S^{+})-i[\mathcal{H},\rho]
\eeq
where $\gamma$ is the relaxation rate, $S^+$ and $S^-$ are the electron spin raising and lowering operators ($S^\pm=S_x \pm iS_y$), and $\beta$ relates the electron Zeeman splitting to $k_B T$, as defined above. This can be simplified by transforming into the rotating frame of the Hamiltonian, taking an Ising approximation ($\mathcal{H}_0=\omega_e S_z - \omega_I I_z + A \!\cdot\! S_z \!\cdot\! I_z$). In this frame, $\mathcal{H}$ goes to 0, and we are left only with the relaxation part of Eq. (7), with $S^+$ and $S^-$ transformed into the rotating frame. Neglecting direct nuclear relaxation (\tonen $\rightarrow \infty$) and in the high temperature limit this yields:

\beq
\dot{\rho}\simeq-\frac{\gamma}{2}
\left(%
\begin{array}{cccc}
\rho_{1,1}(t)-\rho_{2,2}(t) & \rho_{1,2}(t) & \rho_{1,3}(t)-e^{-iAt}\rho_{2,4}(t) & \rho_{1,4}(t)\\
\rho_{2,1}(t) & \rho_{2,2}(t)-\rho_{1,1}(t) & \rho_{2,3}(t) & \rho_{2,4}(t)-e^{iAt}\rho_{1,3}(t)\\
 \rho_{3,1}(t)-e^{iAt}\rho_{4,2}(t) & \rho_{3,2}(t) & \rho_{3,3}(t)-\rho_{4,4}(t) & \rho_{3,4}(t)\\
\rho_{4,1}(t) &  \rho_{4,2}(t)-e^{-iAt}\rho_{3,1}(t)& \rho_{4,3}(t) &  \rho_{4,4}(t)-\rho_{3,3}(t)\\
\end{array}%
\right)
\label{maineq}
\eeq

The electron relaxation rate $(1/T_{1e})$ can be ascertained by observing the appropriate density matrix elements:
$$\dot{\rho}_{1,1}+\dot{\rho}_{3,3} = -\frac{\gamma}{2}(\rho_{1,1}+\rho_{3,3})+\frac{\gamma}{2}(\rho_{2,2}+\rho_{4,4})=-\frac{\gamma}{2}(\rho_{1,1}+\rho_{3,3})+\frac{\gamma}{2}(1-\rho_{1,1}-\rho_{3,3})$$

Taking $\rho_e=\rho_{1,1}+\rho_{3,3}$ then, $\dot{\rho}_e=-\gamma(\rho_e-1/2)$. Solving this gives:
\beq
\rho_e=\rho_{e,0} e^{-\gamma t}+1/2
\eeq

Hence, electron relaxation follows $e^{-\gamma t}$ and the electron relaxation time, $T_{1e}=\frac{1}{\gamma}$.
\\

The nuclear coherence is given by $\rho_{nn}=\rho_{3,1}+\rho_{4,2}$. Extracting these terms from Eq.~\ref{maineq} yields two coupled differential equations:
\beq
\left(
\begin{array}{c}
\dot{\rho_{3,1}}\\
\dot{\rho_{4,2}}
\end{array}\right)
= - \frac{\gamma}{2}
\left(\begin{array}{cc}
1 & -e^{iAt}\\
-e^{-iAt}& 1
\end{array}
\right)
\left(
\begin{array}{c}
\rho_{3,1}\\
\rho_{4,2}
\end{array}\right)
\eeq

It is straightforward to get rid of the time dependence in the $2\times2$ matrix in this equation by making a time dependent unitary transformation and solving for new variables $\rho_{3,1}'$ and $\rho_{4,2}'$.
\beq
\left(\begin{array}{c}
\rho_{3,1}'\\
\rho_{4,2}'
\end{array}\right) = U\left(\begin{array}{c}
\rho_{3,1}\\
\rho_{4,2}
\end{array}\right) = \left(\begin{array}{cc}
e^{-iAt/2}& 0\\
0 & -e^{iAt/2}
\end{array}\right)\left(
\begin{array}{c}
\rho_{3,1}\\
\rho_{4,2}
\end{array}
\right)
\eeq
Following this transformation solving the pair of differential equations is a simple eigenvalue problem. In the experiments $A = 117$~MHz and $\gamma$ ranges from 1~kHz to less than 1~Hz (as a function of temperature), hence we can take the limit $A\gg\gamma$. In this case, both characteristic eigenvalues have a real part of $-\gamma/2$, and therefore any nuclear coherence decays with this rate. Thus $T_{2n}=\frac{2}{\gamma}=2T_{1e} $ as experimentally observed.

\subsection{C. Density matrix tomography}
In this section we describe 1) the preparation and the tomography of the pseudopure initial electron spin states $\pm X$, $\pm Y$, and $\pm Z$, and the Identity, 2) the tomography of the state after transfer to and from the nuclear spin and 3) the measure of fidelity between the starting and recovered states. 

We define state $\pm X$ as ($\pm\sigma_x + \mathbb{I})/2$, and similarly for $Y$ and $Z$. \Fig{fig:pulseall} shows the full set of pulse sequences required for the state preparation and detection. The starting (thermal) state is +Z, which thus requires no preparation pulse. $-Z$ is obtained by applying an inversion $\pi$ pulse, while $\pm X$ and $\pm Y$ are obtained through a $\pi/2$ pulse of the appropriate phase. The state $\mathbb{I}$ is obtained by applying an inversion $\pi$ pulse and waiting some time, $T = (\ln{2}) T_{1e}$. This is long enough to ensure complete decoherence of the electron spin (off diagonal elements go to zero), while corresponding to the precise point during the relaxation process at which the electron spin populations are equal.

Measurement is performed in the $\sigma_x$ and $\sigma_y$ bases by generating an electron spin echo and observing both in-phase and quadrature components. A measurement in the $\sigma_z$ basis can be performed some short time ($t < T_{\rm{1e}}$) later in the pulse sequence by applying a $\pi/2$ pulse, followed by a $\pi$ pulse, and then observing the resulting echo. This measurement operation is applied to both the starting states and those recovered after the end of the \emph{write-read} process.

\begin{figure}[t] \centerline
{\includegraphics[width=7in]{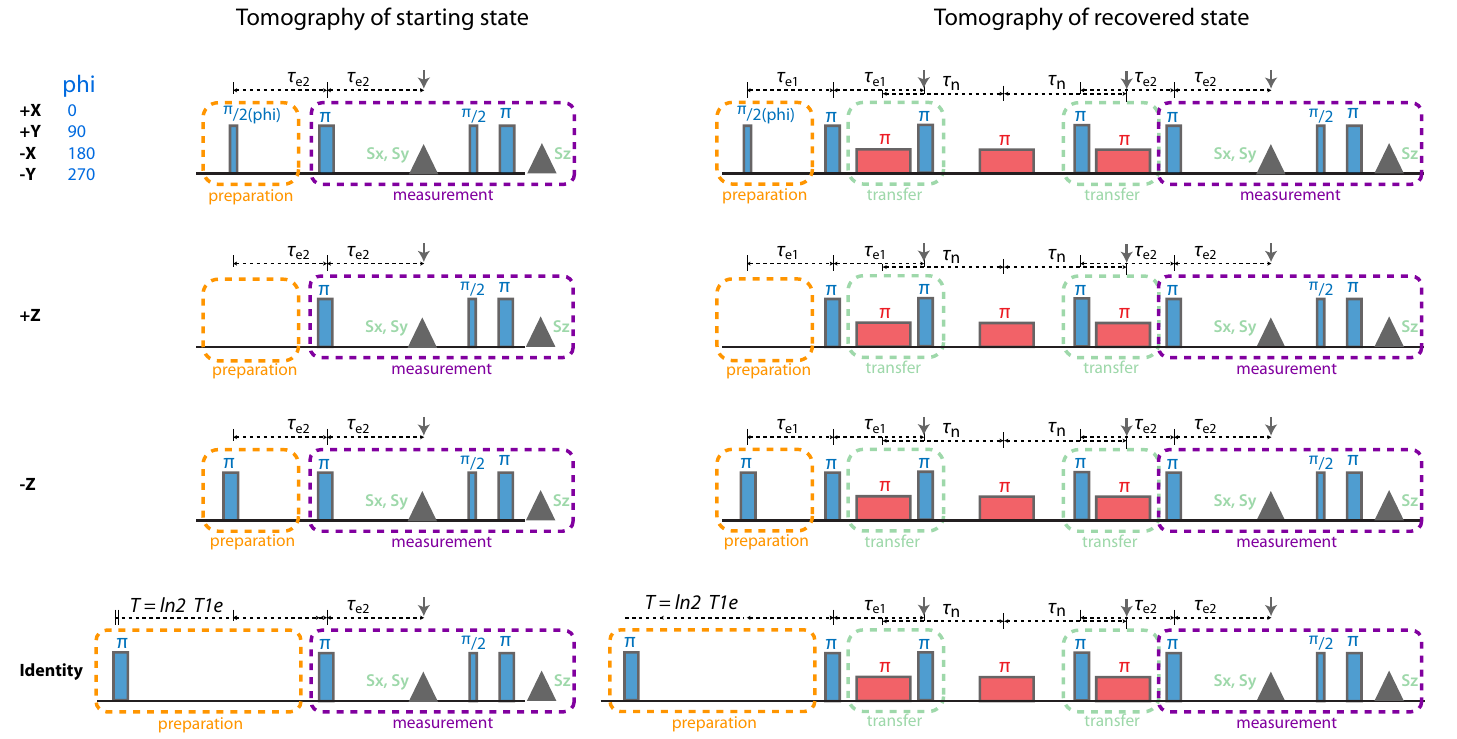}} 
\caption{\textbf{Pulse sequences applied to prepare and measure electron spin states.}} \label{fig:pulseall}
\end{figure}

Echo traces for the seven states are shown in \Fig{fig:allechoes} for both the starting and recovered states. Each corresponds to a measurement in the $\sigma_{x,y,z}$ bases.

\begin{figure}[t] \centerline
{\includegraphics[width=7in]{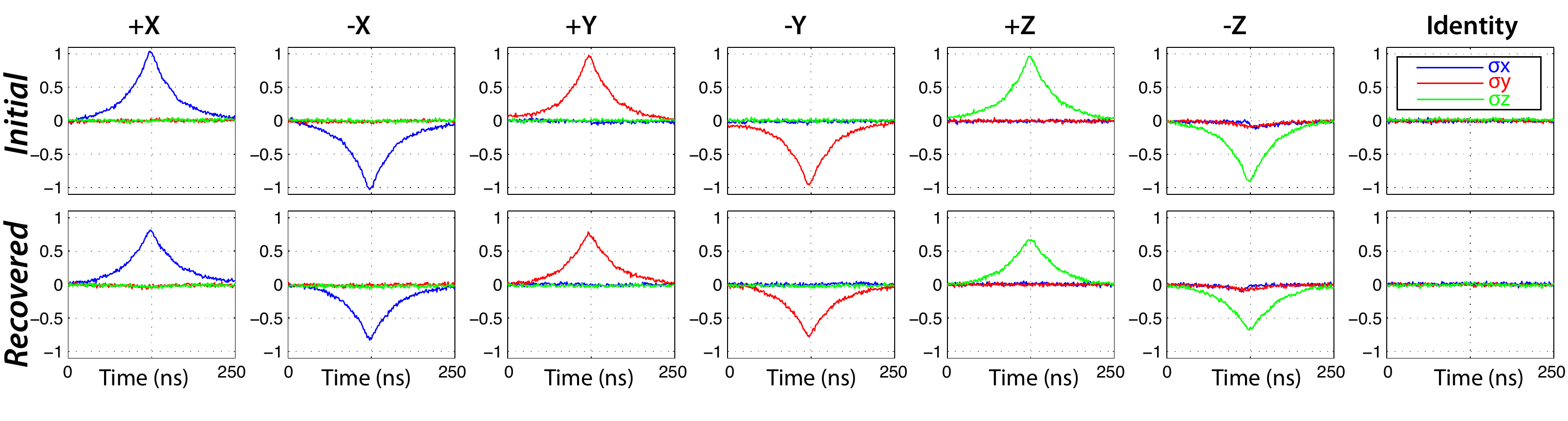}} 
\caption{\textbf{Measurements in the $\sigma_{x,y,z}$ bases of the electron spin state before and after storage in the nuclear spin.}
Electron spin echoes (a.u.) are obtained using the pulse sequences shown in \Fig{fig:pulseall}. The echoes for $\sigma_x$ and $\sigma_y$ occur simultaneously, while that for $\sigma_z$ (which occurs at a later time) is superimposed here for clarity.} \label{fig:allechoes}
\end{figure}

The integrated areas of the electron spin echoes, $A_{x,y,z}$, are used to extract the components of $\sigma_{x,y,z}$ in the spin density matrix. We assume the starting electron spin state is (pseudo)pure and can thus normalise the areas to extract a density matrix of the starting electron spin state:
\beq \rho = \frac{A_x \sigma_x + A_y \sigma_y + A_z \sigma_z}{2\sqrt{A_x^2+A_y^2+A_z^2}} + \mathbb{I}/2 \eeq

We make no such assumptions about the purity of the recovered electron spin state, and normalise the integrated areas of the recovered spin echoes using the areas obtained from the starting state. Density matrices for the initial and recovered state are shown in \Fig{fig:allrhos}.

\begin{figure}[t] \centerline
{\includegraphics[width=7in]{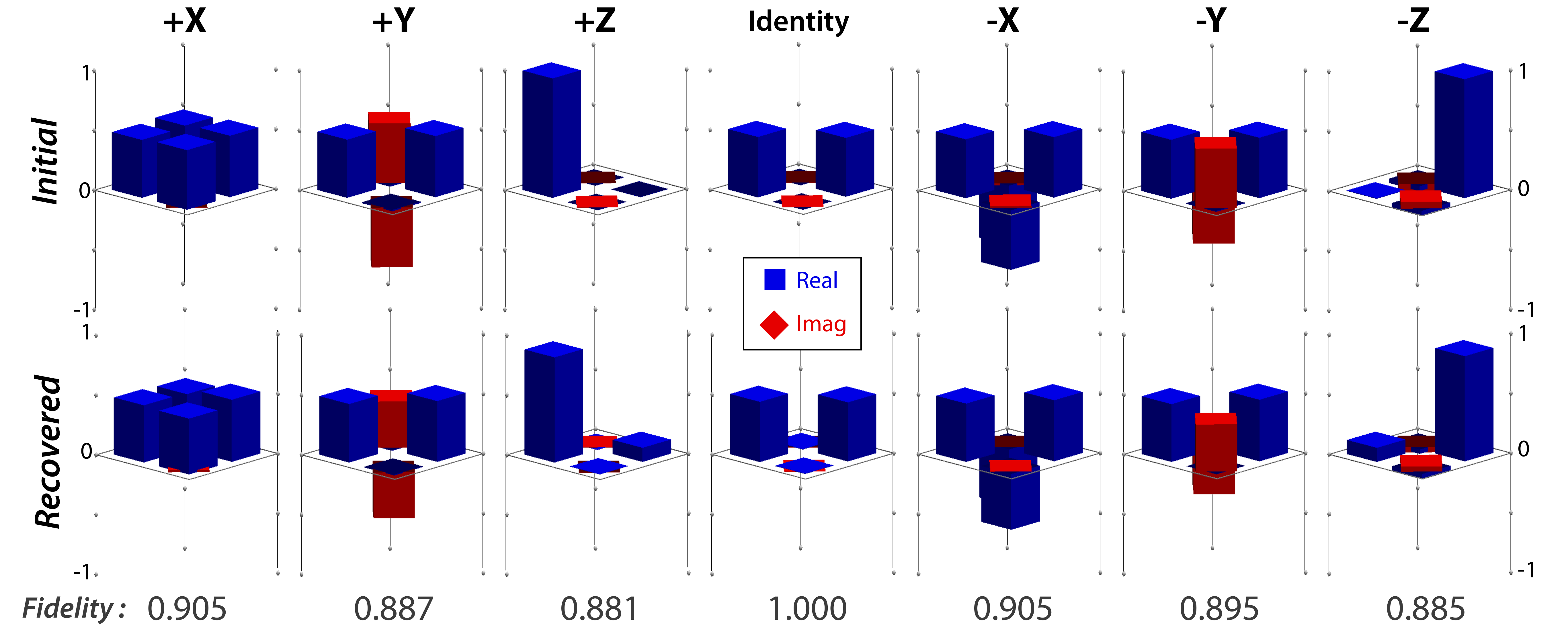}} 
\caption{\textbf{Density matrices for the initial and recovered states.}} \label{fig:allrhos}
\end{figure}

One common measure of the difference between two quantum states is the \emph{fidelity}~\cite{nielsenchuang}:
\beq
F(\rho_0,\rho_1)= \rm{Tr}\left(\sqrt{\sqrt{\rho_1}\rho_0\sqrt{\rho_1}}\right)
\eeq
Here, we use a more aggressive measure of fidelity, $F'=F^2$, corresponding to the overlap of a pure state and an arbitrary density matrix (rather than its square root). Thus, if our initial density matrix $\rho_0 = \ket{\psi}\bra{\psi}$, the fidelity measure we use is:
\beq
F'=\bra{\psi}\rho_1\ket{\psi}.
\eeq

\end{widetext}
\end{document}